 \documentclass[preprint2]{aastex}

\shorttitle{MIR spectroscopy of high--$z$ 3CRR sources}
\shortauthors{Leipski et al.}

\usepackage{multirow}

\begin{document}


\title{Mid-infrared spectroscopy of high-redshift 3CRR sources}


\author{C. Leipski\altaffilmark{1,2}, 
M. Haas\altaffilmark{3},
S.~P. Willner\altaffilmark{4}, 
M.~L.~N. Ashby\altaffilmark{4}, 
B.~J. Wilkes\altaffilmark{4}, 
G.~G. Fazio\altaffilmark{4}, \\
R. Antonucci\altaffilmark{1},
P. Barthel\altaffilmark{5},
R. Chini\altaffilmark{3},
P. Ogle\altaffilmark{7},
R. Siebenmorgen\altaffilmark{6},
F. Heymann\altaffilmark{3,6}
}
\altaffiltext{1}{Department of Physics, University of California,
  Santa Barbara, CA 93106, USA }
\altaffiltext{2}{email: leipski@physics.ucsb.edu}
\altaffiltext{3}{Astronomisches Institut, Ruhr-Universit\"at Bochum,
  Universit\"atsstra{\ss}e 150, 44801 Bochum, Germany }
\altaffiltext{4}{Harvard-Smithsonian Center for Astrophysics, 60 Garden
  Street, Cambridge, MA 02138, USA }
\altaffiltext{5}{Kapteyn Astronomical Institute, University of
  Groningen, P.O. Box 800, 9700 AV Groningen, Netherlands }
\altaffiltext{6}{European Southern Observatory,
  Karl-Schwarzschildstrasse 2, 85748 Garching, Germany} 
\altaffiltext{7}{Spitzer Science Center, California Institute of
  Technology, Mail Code 220-6, Pasadena, CA 91125, USA} 

\begin{abstract}

Using the {\it Spitzer Space Telescope}, we have 
obtained rest frame 9--16~\micron\ spectra of 11 quasars and 9 radio
galaxies from the 3CRR catalog at redshifts $1.0<z<1.4$. 
This complete flux-limited 178\,MHz-selected sample is 
unbiased with respect to orientation and therefore 
suited to study orientation-dependent effects 
in the most powerful active galactic nuclei (AGN).
The mean radio galaxy spectrum shows a clear silicate 
absorption feature ($\tau_{9.7\mu{\rm m}} = 1.1$) whereas the mean 
quasar spectrum 
shows silicates in emission.  The mean radio galaxy spectrum matches
a dust-absorbed 
mean quasar spectrum in both shape and overall flux level.
The data for individual objects conform to these results.
The trend of the silicate depth to increase with decreasing 
core fraction of the radio source further supports that for this sample, 
orientation is the main driver for the difference between radio 
galaxies and quasars, as predicted by AGN unification. 
However, comparing our high-$z$ sample with lower redshift 3CRR
objects reveals that the absorption of the high-$z$ 
radio galaxy MIR continuum is lower than expected from a scaled up
version of lower luminosity sources, and we discuss some effects that may 
explain these trends.

\end{abstract}

\keywords{galaxies: active; galaxies: high-redshift; infrared: galaxies}

\section{Introduction}

There is strong evidence that many luminous active
galactic nuclei (AGN) are surrounded by dust in a torus-like geometry
leading to orientation-dependent obscuration of the nuclear region
\citep[][]{ant85,bar89,ant93,urr95}.  
Despite the general consensus on the AGN unification principle, it is
still a matter of debate whether obscuration is indeed the dominating effect on
the observed differences between radio galaxies and quasars. The
question remains if every  obscured (type-2) AGN has a (type-1) AGN 
hidden by a ``torus'' or if some of the observed differences may arise 
from other effects, for instance different evolutionary states, different star forming 
activity in the hosts, or different accretion rates.

Orientation-dependent effects can only be tested  with
type-1 and type-2 AGN samples matched in isotropic emission.
Low-frequency (meter wavelength) radio-selected AGN are particularly
attractive for studying orientation-dependent properties at other
wavelengths because the integrated emission from the radio lobes is
optically thin and essentially isotropic.  This paper therefore focuses 
on 
powerful double-lobed radio galaxies (type-2) and steep-spectrum quasars
(type-1) taken from the 3CRR catalogue \citep{lai83}, a subset of the 
deeper and larger 3CR sample \citep{spi85}. The 3CRR sample provides a complete, flux-limited 
sample of radio-loud objects selected at 178\,MHz.
Mid infrared (MIR) and far infrared (FIR) observations of orientation-unbiased samples 
are valuable tools for the study of unification schemes. They probe the 
intrinsic energy budget of the 
nuclear regions where the optical/UV radiation from the 
accretion disk is absorbed by the surrounding dust, which then re-emits the absorbed 
energy at infrared wavelengths. Thus, the infrared might serve as a ``calorimeter'' 
for the radiative output in the nucleus. However, while the infrared radiation 
has the ability to penetrate the obscuring dust, it will still suffer some 
(wavelength-dependent) extinction, which may yield important clues as to the nature of the obscurer itself.

At lower redshift ($z<0.5$) the large number of radio galaxies compared to quasars 
is inconsistent with orientation-dependent models 
\citep[e.g., Fig.~1 of][]{sin93}.  It has been argued that this might be due 
to an additional population of low-luminosity radio galaxies 
which  falls below the selection limit in typical higher-redshift, low-frequency-selected samples. 
These low-luminosity radio  galaxies often show quite different multi wavelength properties 
with moderately absorbed (and less luminous) accretion-related X-ray 
emission \citep[e.g.][]{har09}, optical spectra  of low ionization 
\citep[e.g.][]{but09}, and weak MIR emission \citep{mei01,ogl06,vanderWolk2010}.

At intermediate redshifts ($0.5<z<1.0$) numerous studies --- often
utilizing IR data --- show that the unification picture works well for
radio galaxies and quasars \citep[e.g.][]{bar89,haas04}.  To date, a
large number of $z < 1$ 3CR objects have been studied in the MIR and
FIR with {\it ISO} (e.g. \citealt{vanb00};\citealt{mei01}; and as
compiled by \citealt{sieb04} and by \citealt{haas04}) and with {\it
Spitzer} \citep[e.g.][]{shi05,haas05,ogl06,cle07}.
While such studies were vital to establish the unification picture at 
$0.5<z<1.0$, some of this work also indicated that on average quasars 
seem to be a factor of a few more luminous in the infrared than radio 
galaxies when normalized by their total radio luminosity. 
To explain this effect, it was argued that non-thermal contributions of 
a beamed radio core increase the infrared fluxes of quasars in addition 
to the wavelength-dependent obscuration of the torus in radio galaxies 
\citep[e.g.][]{hes95,cle07}.
\citet{mei01} and \citet{haas04} saw the MIR/FIR difference 
between quasars and radio galaxies but did not model it in detail 
(though the simplest explanation --- obscuration --- was briefly mentioned), and 
\citet{sieb04} have interpreted the difference in accordance with an obscuration scenario.

At high redshifts ($z>1.0$) the situation is less clear. For the 3CR 
sources at these redshifts, sensitive FIR (60-200\,$\mu$m) observations 
have yet to be made, but 3.6--24\,\micron\ photometry of the complete 
high-$z$ part ($1.0<z<2.5$) of the 3CR sample has been obtained with 
{\it Spitzer}. The rest frame 2--10\,\micron\ spectral energy distributions 
(SEDs) provide evidence for dust emission which is a factor 3--5 
weaker in type-2 AGN compared to type-1 AGN \citep{haas08}. The 
mean radio galaxy SED is consistent with a reddened version 
of the mean quasar SED, leaving little room for 
a significant contribution of beamed non-thermal emission to the infrared 
SEDs of quasars, at least for that sample.

Apart from the unification issues, it is vital to 
further explore the largely unknown MIR properties of high redshift 
radio-loud AGN and to expand the luminosity range and potentially the 
AGN-host contrast studied so far. 
Therefore, we here analyze {\it Spitzer} spectra of the quasars and 
radio galaxies in the 3CRR sample at redshift $1<z<1.4$.
We use a $\Lambda$CDM cosmology with $H_0
=71$~km~s$^{-1}$~Mpc$^{-1}$, $\Omega_{{\rm m}} = 0.27$, and
$\Omega_{\Lambda} = 0.73$.

\begin{table*}[t!]
\begin{center}
\caption{\protect\centering The Sample.\label{tab1}}
\begin{tabular}{lcccccr@{.}lcr@{.}l}
\tableline\tableline
 Object      &  $z$   & $F_{178\,{\rm MHz}}$ &    $\alpha$       & $F_{5\,{\rm GHz}}^{\rm core}$ & references & \multicolumn{3}{c}{$F_{15\,\mu{\rm m}}$} & \multicolumn{2}{c}{$\tau_{9.7\mu{\rm m}}$}\\ 
  &  & (Jy)                 &   & (mJy)                   & & \multicolumn{3}{c}{(mJy)}   & \multicolumn{2}{c}{}                 \\
\multicolumn{1}{c}{(1)} & (2) & (3) & (4) & (5) & (6) & \multicolumn{2}{c}{(7)} & (8) & \multicolumn{2}{c}{(9)} \\
\tableline
\multicolumn{11}{c}{Quasars}\\	       	    
\tableline			       	    		 
3C\,068.1 & 1.238 & 14.0 & 0.80 &                     1.1  &  4 &  9&3 & 0.7 & -0&05 \\ 
3C\,181   & 1.382 & 15.8 & 1.00 &                     9.0  &  8 &  8&7 & 0.5 &  0&19 \\ 
3C\,186   & 1.063 & 15.4 & 1.15 &                    15.0  &  7 &  8&2 & 0.6 & -0&12 \\ 
3C\,190   & 1.197 & 16.4 & 0.93 &                    73.0  &  7 & 13&4 & 1.5 &  0&60 \\ 
3C\,204   & 1.112 & 11.4 & 1.08 &                    26.9  &  4 &  8&3 & 0.5 & -0&13 \\ 
3C\,208   & 1.109 & 18.3 & 0.96 &                    51.0  &  4 &  5&8 & 0.5 & -0&27 \\ 
3C\,212   & 1.049 & 16.5 & 0.92 &                   186.0  &  1 & 15&5 & 0.9 & -0&11 \\ 
3C\,245   & 1.029 & 15.7 & 0.78 &                  1219.0  & 12 & 28&3 & 1.9 & -0&02 \\ 
3C\,268.4 & 1.400 & 11.2 & 0.80 &                    45.0  & 11 & 15&6 & 0.8 &  0&05 \\ 
3C\,287   & 1.055 & 17.8 & 0.42 &                  3280.0  &  9 &  7&1 & 0.6 & -0&23 \\ 
3C\,325   & 1.135 & 17.0 & 0.70 &                     2.4  &  5 &  4&5 & 0.3 & -0&21 \\ 
\tableline
\multicolumn{11}{c}{Radio Galaxies}\\
\tableline
3C\,013   & 1.351 & 13.1 & 0.93 &                     0.18 &  2 &  5&6 & 0.9 &  0&85 \\ 
3C\,065   & 1.176 & 16.6 & 0.75 &                     0.52 &  2 &  2&2 & 0.2 &  0&30 \\ 
3C\,252   & 1.105 & 12.0 & 1.03 &                     2.2  &  5 & 11&8 & 1.0 &  0&61 \\ 
3C\,266   & 1.272 & 12.1 & 1.01 &                     0.4  &  6 &  2&9 & 0.4 &  0&72 \\ 
3C\,267   & 1.144 & 15.9 & 0.93 &                     1.87 & 10 &  6&0 & 0.5 &  0&77 \\ 
3C\,324   & 1.206 & 17.2 & 0.90 &                     0.17 &  3 &  8&3 & 0.7 &  1&71 \\ 
3C\,356   & 1.079 & 12.3 & 1.02 &                     1.1  &  5 &  7&4 & 0.6 &  0&91 \\ 
3C\,368   & 1.132 & 15.0 & 1.24 &                     0.22 &  3 &  8&1 & 0.7 &  1&74 \\ 
3C\,469.1 & 1.336 & 12.1 & 0.96 &                     2.4  & 10 &  8&3 & 1.3 &  1&74 \\ 
\tableline			       	    		 
\end{tabular}
\tablecomments{(1) Object name; (2) Redshift; (3) Observed 178\,MHz flux in Jy; 
(4) Spectral index ($S_{\nu}\propto\nu^{-\alpha}$) between (observed) 178 and 750\,MHz 
as taken from the 3CRR web page; (5) Flux density of the radio core in mJy; (6) References 
for the radio-core flux densities. They refer to observed frame 5\,GHz  except for 
data from \citet{best97}, which were taken at 8.4\,GHz; 
(7) Flux density measured at rest-frame 15\,$\mu$m in mJy; (8)
1$\sigma$ uncertainties on the 15\,$\mu$m 
flux density; (9) Screen opacity relative to 
the mean quasar spectrum as determined in \S\ref{sec:sil}.}
\tablerefs{
   (1) \citealt{aku91};
   (2) \citealt{best97};
   (3) \citealt{best98};
   (4) \citealt{brid94};
   (5) \citealt{fer97};
   (6) \citealt{liu92};
   (7) \citealt{lud98};
   (8) \citealt{man92};
   (9) \citealt{pea85};
  (10) \citealt{ped89};
  (11) \citealt{rei95};
  (12) \citealt{sai90}
}

\end{center}
\end{table*}

\begin{figure*}[t!]
\centering
\includegraphics[angle=0,scale=0.7]{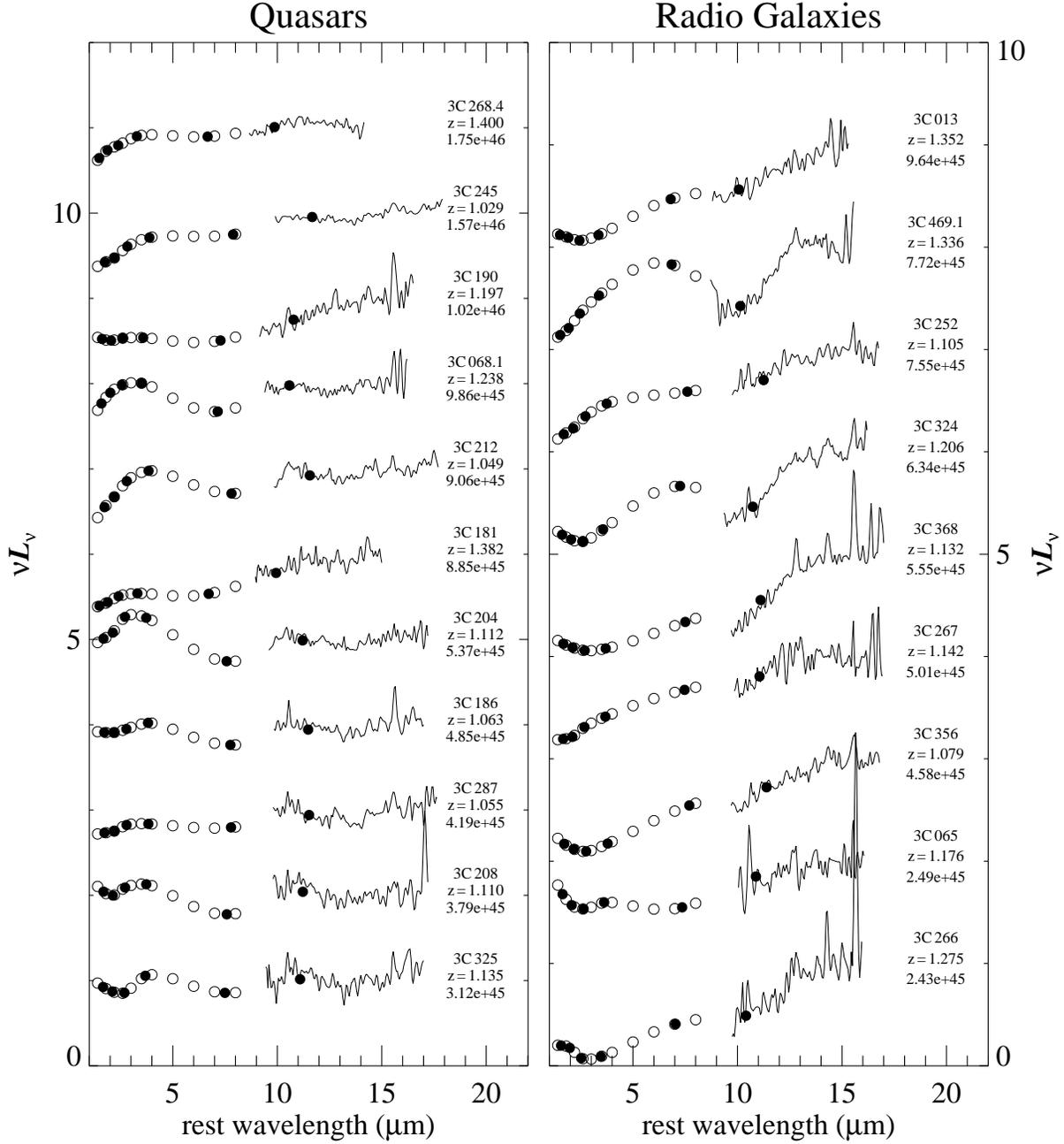}
\caption{Spectra and photometry for the individual objects in the 
rest frame of the sources in arbitrary linear units of $\nu L_{\nu}$.
SEDs have been offset vertically for clarity, but 15\,\micron\ flux
densities can be found in Table~1.  Quasars are shown on the left and 
radio galaxies on the right. The sources are ordered according 
to their 15\,$\mu$m luminosities with $\nu L_{\nu,15\,\mu{\rm m}}$ increasing 
from bottom to the top. The filled symbols show the observed photometry while 
the open symbols represent the interpolated photometry used for 
constructing the average SEDs (Fig.~\ref{fig1}). To the right of each SED
we give the object name, the redshift, and the rest frame luminosity 
 $\nu L_{\nu,15\,\mu{\rm m}}$ in erg\,s$^{-1}$.\label{fig7}}
\end{figure*}

\begin{figure*}[t!]
\centering
\includegraphics[angle=0,scale=.66]{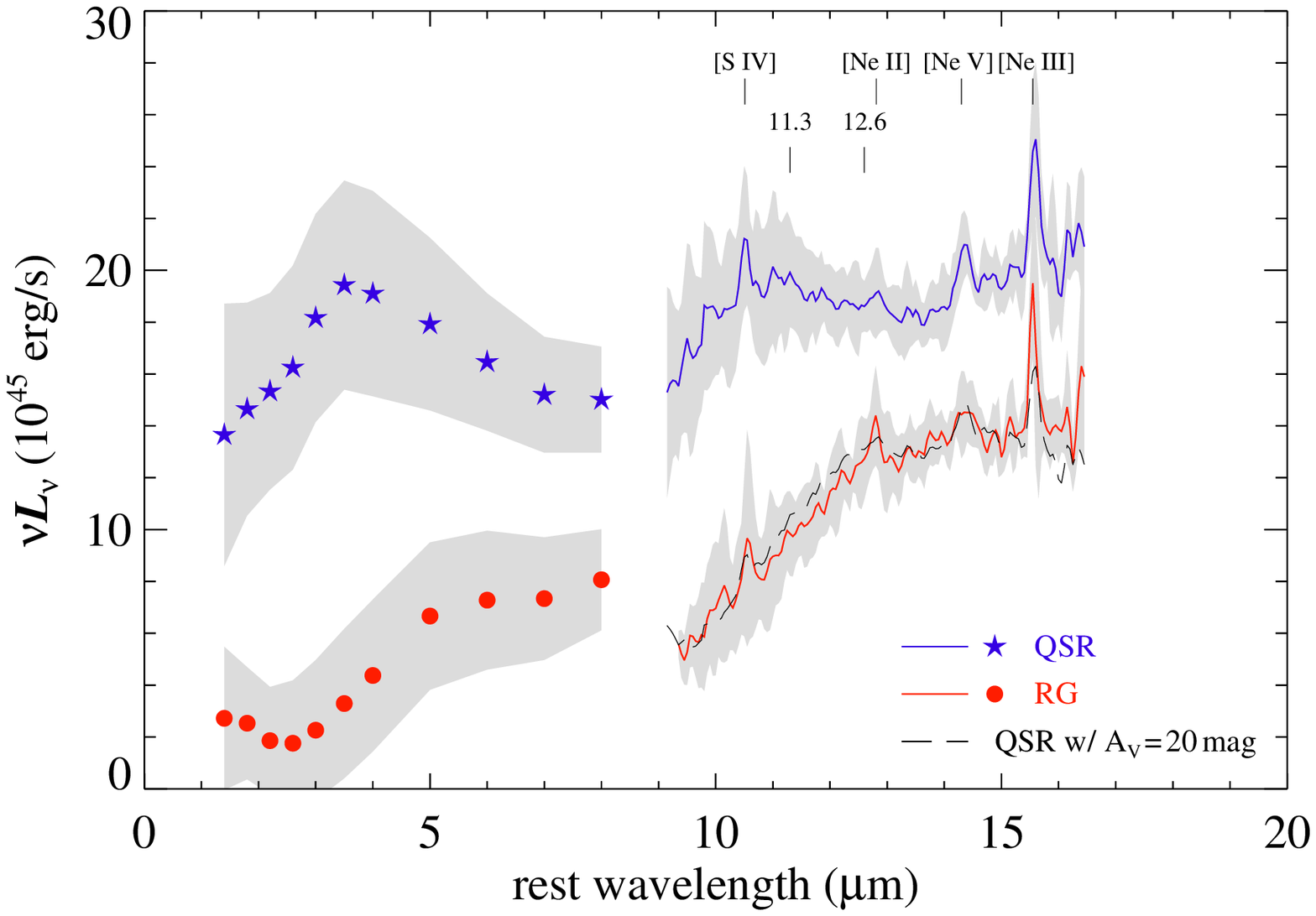}
\caption{Average $9-16\,\mu$m spectra of the high-z 3CRR sources
  supplemented by $1-8\,\mu$m photometry. 
 The solid lines represent the average spectra while the 
 symbols 
  indicate the average photometry. The SEDs are 
  normalized to the isotropic low-frequency radio luminosity. 
  The black dashed line represents the average quasar spectrum reddened 
  by a dust screen of 
  A$_{V}=20$\,mag 
  without any additional scaling. 
  The shaded areas indicate the $1\sigma$ dispersion.
  \label{fig1} }
\end{figure*}

\section{Data}

Our sample comprises all objects of the 3CRR
catalog\footnote{http://3crr.extragalactic.info/} with
$1<z<1.4$\footnote{Due to an updated 
redshift \citep{gri05}, 3C\,325 is included. 4C\,16.49 \citep[$z=1.296$:][]{aars05}
should in principle be part of the sample,  but 
at the time the observations were defined the NED redshift 
was $z=1.88$, and therefore no data were obtained for this source. }. 
This selection includes 20 sources in total, 11 quasars
and 9 radio galaxies. Some basic properties of the sources are
presented in Table~\ref{tab1}.
The redshift range $1<z<1.4$ was selected to cover the rest frame
wavelengths 9--16\,$\mu$m when using the LL1 spectral module of the
Infrared Spectrograph (IRS, \citealt{hou04}) on board {\it Spitzer
Space Telescope} \citep{wer04}. For $1<z<1.4$, the spectral window of
the IRS/LL1  ($19.5-36.5$\,$\mu$m) includes redshifted low- and
high-excitation atomic emission lines 
[\ion{Ne}{2}]\,$\lambda$12.8\,$\mu$m,
[\ion{Ne}{5}]\,$\lambda$14.3\,$\mu$m, and
[\ion{Ne}{3}]\,$\lambda$15.5\,$\mu$m. It also covers potential 11.3\,$\mu$m PAH
emission from star formation and the silicate feature at 9.7\,$\mu$m.

18 objects were observed for a total of 2880 seconds each with IRS in
the LL1 module (PID-40314, PI M.~Haas). 
The remaining objects,
3C\,356 (PID 03349, PI R.~Siebenmorgen) as well as 3C\,325 (PID 00074,
PI J.~Houck), were observed previously in LL1 for a total of
960\,s each.  The individual frames were averaged for each nod
position separately, and the resulting images were used to subtract the
background from the other nod.  The final background-subtracted
images were cleaned of residual rogue pixels and cosmic rays using
{\sc irsclean}. The {\sc spice} software was used to calibrate and
extract one dimensional spectra. The extracted spectra from the two
nod positions were then averaged into a single spectrum.

The comparison of our spectroscopy with photometric measurements in
the MIPS \citep{rie04} 
24\,$\mu$m filter reveals a good agreement
except for 3C\,13, 3C\,65, 3C\,68.1, and 3C\,267, all of which 
show a flux
deficit in the spectra. Checking the pointing for the observation reveals that for 
3C\,13, the spectroscopic slit fell slightly off-center thus missing part of the flux. 
We applied a multiplicative
factor of 1.8 to the spectrum to bring it to the level of the
photometric measurement. For the radio galaxy 3C\,267, the MIPS 24\,$\mu$m image
reveals a faint source to the north-west which is also partly covered by the 
slit. The distance of this faint source from the radio galaxy is close to the 
nod throw along the slit. While this configuration can potentially lead to an 
overcorrection when the two nod positions are subtracted from each other to 
remove the background, we cannot identify the faint source on averaged 
2-dimensional spectral frames at the individual nod positions.  A multiplicative 
factor of 1.2 is needed to match the spectrum with the photometry.
For the other two sources, 3C\,65 and 3C\,68.1,
no immediate cause for the flux difference could be identified.  For
these objects we applied correction factors of 1.5 and 1.3,
respectively.  These corrections do not affect the basic conclusions
drawn here, but we mention that wavelength-dependent slit losses 
 could result in overestimating the silicate depth in 3C\,13 (section \ref{sec:sil}).

\subsection{Creating Average Spectra and SEDs}

All objects have been securely detected, and continuum features such as
silicates in emission or absorption can be observed in individual
sources  (see Table \ref{tab1} for measured rest-frame MIR fluxes and
Fig.\ref{fig7}  for the observed spectra). Because the quality of the
spectra is often limited,  we created separate average spectra for the
radio galaxies and quasars.  First, the individual spectra were
shifted into the rest frame of the source, interpolated onto a common
wavelength grid, and then the interpolated spectra were transformed
into $\nu L_{\nu}$. To avoid the average spectra being dominated by
the most MIR  luminous objects, the spectra were normalized in $\nu
L_{\nu,15\,\mu{\rm m}}$ before averaging.  While this normalization is
well suited to emphasize differences in the spectral shape,
information about the absolute MIR luminosity is removed from the
average spectra. This information, however, may be recovered 
using the extended radio luminosity of the sources.

Due to their selection, the quasars and radio galaxies in this sample
span the same range of radio luminosities. The actual rest frame mean 
values are
$\langle\nu{\rm L}_{\nu,178\,{\rm MHz}}\rangle = (1.9 \pm 0.6) \times
10^{44}$\,erg/s for the quasars and $\langle\nu{\rm L}_{\nu,178\,{\rm
MHz}}\rangle = (2.0 \pm 0.4) \times 10^{44}$\,erg/s  for the radio
galaxies. First we used the 
15\,$\mu$m normalization factors to scale the individual 
$\nu L_{\nu,178\,{\rm MHz}}$ values. Then we normalized the average 
IR SEDs of each class to have the same average $\nu L_{\nu,178\,{\rm MHz}}$  
of $\nu{\rm L}_{\nu,178\,{\rm MHz}} = 2.0 \times 10^{44}$\,erg/s,  
which is the mean rest frame 178\,MHz radio luminosity of the 
complete sample.

In addition to the spectra, observed photometry at 3.6, 4.5, 5.8, 8.0,
16,  and 24\,$\mu$m \citep{haas08} was used to construct average
near-infrared (NIR)/MIR SEDs for the sample. The photometric
measurements were shifted  in wavelength into the rest frame of the
individual sources and then spline  interpolated onto a common
wavelength grid. Figure~\ref{fig7} shows the observed as well as the interpolated
photometry  together with the spectra. After
transforming the interpolated  values into $\nu L_{\nu}$ we applied
the same scaling factors as for the spectra  and then averaged the
photometry.  After normalization, all the quasar SEDs are very similar
to each other in shape (except for the strength of a NIR bump, see
section \ref{sec:avg}), while the radio galaxy SEDs tend to be more
diverse (Fig.~\ref{fig7}). This has already been explored by
\citet{haas08} for their larger and fainter sample  (which includes
the current sample), and the  trends are similar. Despite the diversity in
the radio galaxy SEDs, we did not exclude any object from the averaging
process.

\section{Results and Discussion}

\subsection{Average SEDs}
\label{sec:avg}

Figure~\ref{fig1} shows the average spectra and photometry 
for the radio galaxies and quasars. The combination of the deep
silicate absorption at 9.7\,$\mu$m and the overall lower flux levels
indicate that the radio galaxies on average suffer considerable absorption
throughout the entire MIR continuum relative to quasars.
Applying a reddening screen of $A_V=20$\,mag to
the average quasar spectrum yields a very good match with the radio
galaxy spectrum in the continuum level as well as in the shape of the
silicate absorption feature.\footnote{We here use the MIR extinction curve from 
\citet{chiar06}  normalized to the peak of the silicate feature at 9.7\,$\mu$m  
and $A_{V}/\tau_{9.7\mu{\rm m}} = 18$.}
From the MIR point of view, this strongly supports the idea
that the quasars and radio galaxies in our sample are in fact {\it intrinsically
identical} with the former being essentially unobscured while the
latter suffer from extinction.
The good match of the average
radio galaxy spectrum to the reddened quasar spectrum also implies
that the intrinsic radio galaxy spectrum is not flat. In particular, the
average radio galaxy must have silicate emission underlying the absorption
feature (see section \ref{sec:sil}).

We here assumed a screen scenario where
the absorbing material is placed in front of the emitter.  However, in
the clumpy torus models which in recent years were able to
successfully explain the dust emission in local Seyfert galaxies
\citep[e.g.][]{hoe06,eli08,nen08}, the MIR-emitting zone is not
located at a single radius but is spread over a range of radii,
thus essentially creating a mixed-case scenario for parts of the
absorber. In general, screen extinction
underestimates column density
by different amounts at different
wavelengths.
In addition, the bulk of the MIR emission
will originate at moderate depth within the obscuring structure thus
suffering (far) less extinction than radiation from the accretion disk or from
hot dust located near the accretion disk.  Thus,
depending on the complexity of the geometry, 
the mean $A_{V}$
determined from the spectroscopy represents a lower limit 
and will in most cases underestimate  the total column density
by a large factor (see e.g. \citealt{lev07}).

A notable feature in the average quasar SED is the prominent NIR
bump at $\sim$2--5\,\micron\ (Fig.~\ref{fig1}). This component is, 
to different degrees, present in the SEDs of essentially all individual 
quasars (Fig.~\ref{fig7}). Such NIR bumps have been previously reported for individual 
type-1 AGN \citep[e.g.][]{bar87,rod06,rif09} and can be seen also in larger 
samples of quasars and in composite SEDs \citep[e.g.][]{ede86,elv94,gal07}. 
They are generally identified with graphite dust near the sublimation temperature 
located very close to the active nucleus \citep[e.g.][]{bar87,mor09}.
There is some indication that the strength of this feature increases with increasing 
visible/UV luminosity of the quasar \citep{ede86,gal07}.

\begin{figure}[]
\centering
\includegraphics[angle=0,scale=.54]{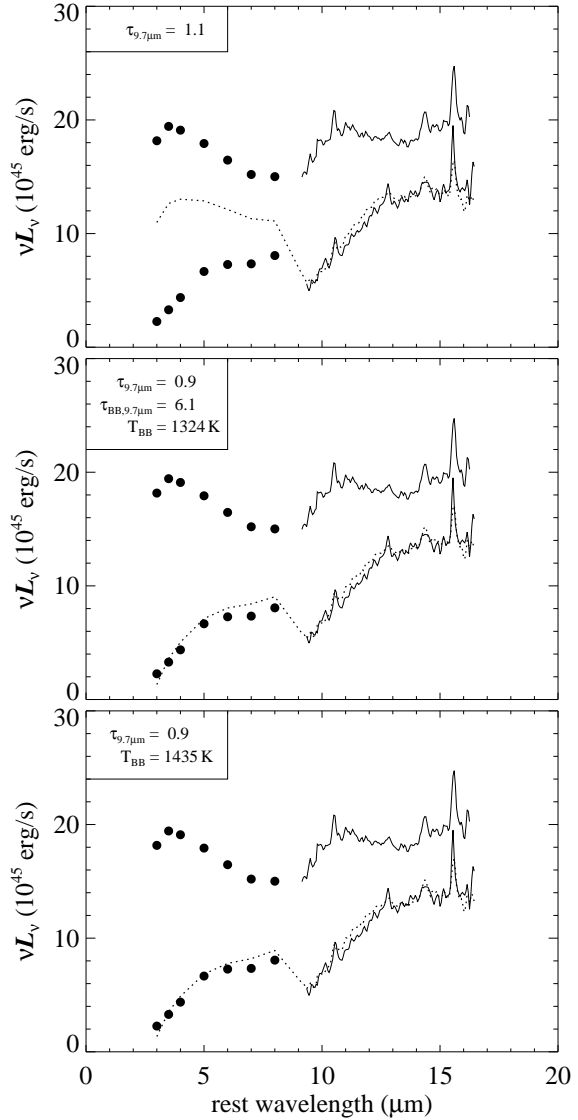}
\caption{Average quasar and radio galaxy SEDs compared to dust screen
  reddening (dashed lines).
No additional scaling was applied. The data are 
those shown in Fig.~\ref{fig1} but  with uncertainties omitted for 
clarity. 
{\it Top}: Quasar SED  reddened using a single $\tau_{9.7\mu{\rm m}}$ value. 
{\it Middle}: Quasar SED reddened with two components, one affecting
all wavelengths and the other, larger one affecting only the single
blackbody component representing the 3--5\,\micron\ bump.
{\it Bottom}: Quasar SED with the blackbody component representing
the 3--5\,\micron\ bump subtracted and the remainder reddened by a
single screen value.\label{avgfit}}
\end{figure}

In contrast to the quasars,  the radio galaxies show no noticeable
2--5\,\micron\ bump.   Screen reddening the 
average quasar SED with $A_V=20$\,mag  leaves an excess for
$\lambda < 8$\,$\mu$m and  
especially at the location of the NIR bump (Fig.\ref{avgfit}, {\it top}). 
This feature is either missing in the radio galaxies, or  it might suffer higher 
extinction than the longer-wavelength continuum.

For their larger sample, \citet{haas08} found that  radio 
galaxy SEDs between 2 and 10\,$\mu$m 
are on average consistent with reddened quasars, similar to our 
result from the spectroscopy at 10--16\,$\mu$m. Because their sample of 3CR 
objects contains a large number of fainter and higher-redshift sources compared 
to the sample of 3CRR objects studied here, we performed the same analysis as 
described by Haas et al.\ but limiting the objects to our 20 sources. Using their 
approach, it would require a screen extinction of 
$A_V\sim 40$\,mag to match the average 2--10\,$\mu$m SEDs of radio 
galaxies and quasars in our sample. This value is higher than the value 
determined from the MIR spectroscopy alone ($A_V\sim\,20$\,mag, 
$\tau_{9.7\mu{\rm m}}\sim1.1$\,mag) as discussed above.

For the discrepancy at $\lambda \leq 8$\,$\mu$m,  several
possibilities can be imagined.  One example is inaccuracies in the MIR
extinction  curves. Different extinction curves show  differences in
the continuum at $\sim$8\,$\mu$m  and in the silicate feature
\citep[e.g.][]{mat83,rie85,chiar06}, and  an incorrect ratio between
continuum and silicate extinction could be part of the
problem. However, considering the good fit in the silicate depth and
in the continuum extinction  for the spectra (Fig.~\ref{avgfit}, {\it
top}),  the extinction curve used here  \citep{chiar06}
works well at least at longer wavelengths. Moreover, published extinction
curves \citep[e.g.,][]{ind05,Nishiyama2009} are all similar and equally 
featureless shortwards of the silicate
feature.  With curves such as these, it is not
possible to match the NIR and MIR 
simultaneously in a screen  configuration affecting all wavelengths
equally.

An alternative solution can be found by including in the fits
a separate blackbody component representing the NIR bump. 
Although a solid theoretical foundation for the origin and the 
properties of the NIR bump is still missing, 
it can be well described as a single temperature 
blackbody \citep[e.g.][]{bar87,mor09}. If the bump comes from dust,
the necessary high temperature places this
dust component close to the nucleus, and the
radiation from this component would have
to penetrate a larger dust column than the rest of the continuum emission. 
Assuming
a black body component from nuclear hot dust suffers more extinction
than the rest of the IR continuum, screen-reddening of the average
quasar SED is consistent with the average radio galaxy SED
(Fig.~\ref{avgfit}, {\it middle}), and the derived
temperature of the NIR bump blackbody ($T\sim1300$\,K) is in the
range expected from previous studies in
the literature \citep[e.g.][]{mor09}.   The major requirement in this scenario is that
the screen opacity for the NIR bump component ($\tau_{9.7\mu{\rm
    m}}\approx6.1$) must be significantly higher than  the 
opacity affecting other components ($\tau_{9.7\mu{\rm m}}\approx0.9$).

It could also be suggested that the NIR bump is absent in most
radio galaxies. Subtracting a single-temperature blackbody
from the average quasar spectrum before  screen reddening the
remaining continuum gives a good fit as well (Fig.~\ref{avgfit}, {\it
bottom}). However, this scenario would imply that there are intrinsic
differences between  radio galaxies and quasars, where the latter host
an additional dust component close to  the nucleus.  Such a scenario
seems unlikely for the majority of the objects. 
Visible-light imaging and polarization studies  show that
at  $0.7 \lesssim z \lesssim 2.5$, essentially all luminous radio
galaxies show the so-called  alignment effect \citep{cham87,mcca87},
and much of this emission turns out to be scattered  light from a hidden
quasar \citep[e.g.][]{cim93,cim97,ver01}.  Thus, at higher redshifts
most if not all radio galaxies host a hidden quasar, arguing for
the similarity of both  types of objects. This also strongly argues 
that the main difference between these two types of objects is 
in the orientation  of their nuclear structures with respect to us. Further
clues on  the nature of the NIR bump might be provided by detailed
radiative transfer modeling of the SEDs  of the sources (F.\ Heymann,
in preparation).   Regardless of the reason  why  the NIR bump is  not
seen in radio galaxies, the estimated  value for $\tau_{9.7\mu{\rm
m}}$ changes very little. That is because it is determined  mostly by
the (rest frame) 10--13\,$\mu$m slope and secondarily by the (rest
frame) 8\,$\mu$m photometry and hardly at all by shorter-wavelength
assumptions  or data. The rest of this paper uses only
$\tau_{9.7\mu{\rm m}}$, and the NIR difference  does not affect our
main conclusions.

\subsection{Individual Spectra}
\label{sec:sil}

\subsubsection{Fitting the Silicate Feature}

A striking feature in Figure~\ref{fig1} is the deep 9.7\,$\mu$m
silicate  absorption observed for the radio galaxies.  The long
wavelength side of this feature is  well measured in the
spectra, and the shorter wavelength photometry data confirm the
presence and depth of this feature.  The silicates alone provide clear
evidence that in the mean radio galaxy, a  substantial amount of
absorbing dust lies between the MIR continuum emitter  and the
observer.

For further analysis we determined the depth of the silicate feature
in the  individual objects by fitting the observed spectrum and
photometry with an input  spectrum reddened by a dust
screen. Considering the uncertainties in  treating the NIR bump
component, we limited the fits for the individual objects  to $\lambda
\geq 7$\,$\mu$m.  We again used the extinction curve presented by
\citet{chiar06} and assumed that  the intrinsic continuum of every
source resembles the average quasar SED,  including silicates in
emission.  In order  to parameterize the average quasar SED, it was
fit by a linear combination of  three blackbodies of variable
temperature. One of these blackbodies  was multiplied by the MIR
extinction curve to include a silicate emission feature 
We do not assign any  physical  interpretation to this fit as it was only used
to yield a smooth representation of the  average quasar
continuum. This smooth continuum was then taken as the intrinsic
continuum for the reddening  fits to the individual objects.

\begin{figure}[t!]
\centering
\includegraphics[angle=0,scale=.38]{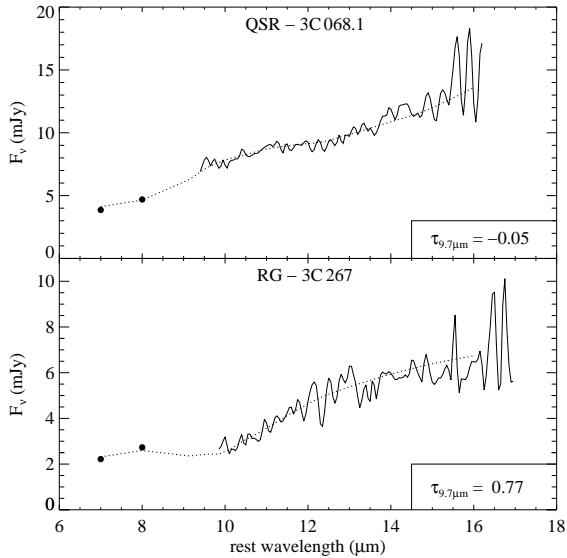}
\caption{Two examples illustrating the fitting of the silicate feature for a quasar ({\it top}) and a radio galaxy ({\it bottom}). The solid line and filled circles show the spectroscopy and interpolated photometry, respectively. The dashed line shows the fit which is a scaled and screen reddened version of the average quasar SED. \label{figfit}}
\end{figure}

Figure~\ref{figfit} shows two examples for our fitting procedure, and
the  derived values for $\tau_{9.7\mu{\rm m}}$ for all objects 
are given in Table~\ref{tab1}.  As in the
radio galaxy average spectrum, all individual galaxy data longwards  of
$\sim$3\,$\mu$m can be fitted by introducing stronger reddening of
a  NIR bump (or assuming such a component is absent). In  the majority
of cases, variations of the 
screen $\tau_{9.7\mu{\rm m}}$ with other fit parameters
are  smaller than $\pm$0.1. Therefore
we assign this value as an uncertainty to  the individual
$\tau_{9.7\mu{\rm m}}$ screen values.

Silicates appear in absorption in all radio galaxies and, with the
exception of 3C\,190, in emission in all quasars. Thus the average
spectra  accurately represent the two  populations.
Unlike
Galactic sources, however, the silicate emission feature  in the
average quasar spectrum peaks at $\sim$11\,$\mu$m.  Such a shift in
the peak wavelength  of the silicate emission is frequently observed
in AGNs \citep[e.g.][]{sieb05,stu05,hao05,shi06}. Under the assumption
that the  emission is optically thin, the shift can be explained as a
result of  folding the silicate opacity curve with the (steeply
rising) black body curve of the silicate grains \citep{sieb05}. Lower
temperatures  would then shift the peak of the silicate feature to
longer wavelengths.  Alternative explanations for the shift of the
silicate emission peak wavelength  in AGN include variations in the
dust grain size  distribution \citep[e.g.][]{stu05}, effects from dust
chemistry  \citep[e.g.][]{mar07,li08}, or radiative transfer effects
\citep[e.g.][]{nik09}.

The only quasar not showing silicate emission is 3C\,190, which shows
a shallow silicate absorption feature instead (Table~1).   
3C\,190 also has the reddest MIR
spectrum of all quasars studied in this paper. Notably, 3C\,190 is a
compact steep spectrum (CSS: projected  linear radio sizes smaller than
15\,kpc) source in the radio, a class of objects which  may represent
young or frustrated sources. This class may show different MIR properties
compared to large double-lobed sources (P.\ Ogle, in
preparation). However, the two other CSS sources in our sample 
(the quasars 3C\,186 and 3C\,287) appear  very similar to the other quasars
in their MIR properties, including obvious silicate emission.

The radio galaxy 3C\,65 stands out by having a very small
$\tau_{9.7\mu{\rm m}}$ value compared to the other radio galaxies.   
It was also one of the sources where
we applied a multiplicative factor to match  the spectrum with the
higher photometric 24\,$\mu$m flux. As discussed in \S2,
pointing errors are more likely to cause an overestimate than an underestimate of
$\tau_{9.7\mu{\rm m}}$.

\subsubsection{Silicate feature and orientation}

\begin{figure}[t!]
\includegraphics[angle=0,scale=.46]{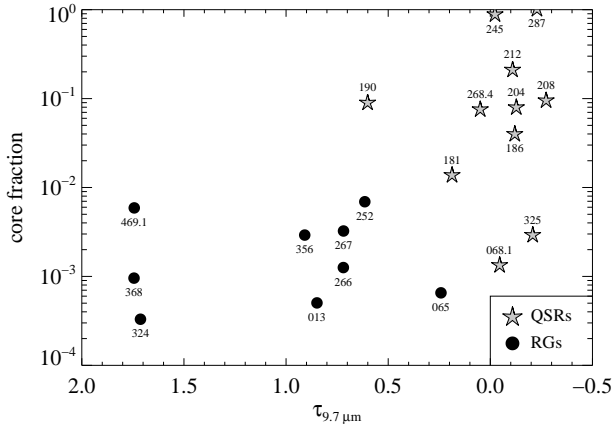}
\caption{Radio core fraction plotted versus 
the depth of the silicate feature. The core fraction was determined 
as the ratio between the 5\,GHz core flux and the 5\,GHz total 
flux.\label{fig5}}
\end{figure}

The observed extinctions can be connected with the orientations
of the sources by analyzing the radio properties of the objects
(Table~\ref{tab1}).  Due to relativistic beaming effects, the
5\,GHz flux of the compact radio core in our objects depends to some extent
on the orientation of the radio jet. Therefore the ratio of the radio core
flux and the (isotropic) extended flux can be used as an indicator of
the approximate inclination of the radio source
\citep[e.g.,][]{orr82}. Because the radio jet is thought to be oriented
along the polar direction with respect to the obscuring structure, one
might expect an anti-correlation between the core fraction and the
extinction in the optical and MIR.\footnote{We here determine the core
fraction as the ratio between the 5\,GHz radio core flux 
(Tab.~\ref{tab1}) and the 5\,GHz total flux taken from \citet{lai80}. 
We used the observed fluxes and did not apply a K-correction.}
This has been demonstrated to be the case from
the optical spectra of quasars which show redder continua (and larger
Balmer line ratios) with decreasing core fraction
\citep[e.g.][]{bak95}. Figure~\ref{fig5} shows the core
fraction of our objects as a function of $\tau_{9.7\mu{\rm m}}$.  All 
sources in this figure with high silicate extinction have weak cores, 
while most sources with silicate emission show stronger cores. This supports 
an orientation-based explanation for the differences between the MIR properties 
of radio galaxies and quasars in the current sample.

\subsubsection{Silicate extinction and 15\,$\mu$m luminosity}

\begin{figure}[t!]
\includegraphics[angle=0,scale=.46]{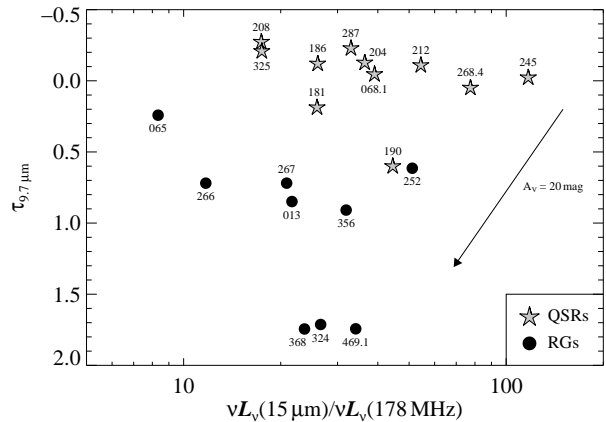}
\caption{Low-frequency normalized luminosity at 15\,$\mu$m plotted versus the depth of 
the silicate feature. The arrow shows the effect of a dust screen with $A_V=20$\,mag, 
assuming the \citet{chiar06} extinction curve. 
Dereddening the radio galaxies along the direction of the $A_V$
arrow shifts them to the same normalized $L_{15 \mu m}$-range as populated by the quasars. 
\label{fig4}}
\end{figure}

While earlier studies \citep[e.g.][]{haas08} showed that the 
MIR difference between quasars and radio galaxies is consistent 
with absorption, the actual amount of absorption was still unknown.
This necessary piece of information can be derived from the silicate
feature depth.
The average spectra in Figure~\ref{fig1} already indicate that radio galaxies
are {\it on average} less MIR luminous than quasars and that this is
consistent with absorption.
Figure~\ref{fig4} shows the comparison for individual
objects. Quasars and radio galaxies show  considerable overlap 
in their 15\,$\mu$m luminosities, but
the depth of the silicate feature separates the two classes almost completely.
Radio galaxies occupy a larger range 
of observed $\tau_{9.7\mu{\rm m}}$ than quasars but with
systematically higher values.
The striking feature of Figure~\ref{fig4} is that dereddening will 
match up the 
radio galaxies with the quasars in both the silicate strength 
and radio-normalized MIR luminosity. 

The characterization of radio galaxies as reddened quasars
appears to be in conflict with results at lower redshifts.
For objects with strong radio cores, \citet{hes95} and \citet{cle07} have 
proposed that beamed synchrotron emission can make significant contributions 
even at MIR wavelengths. 
\citet{cle07} compared  the MIR
continuum properties of a sample of $0.5<z<1.0$ radio galaxies and quasars  also
selected from the 3CRR catalogue. They found that, normalized  to
their 178\,MHz luminosities, quasars are on average $\sim$4 times
more luminous than radio galaxies at (rest) 15\,$\mu$m. 
Based on SED fitting with a spherically symmetric dust model, 
\citeauthor{cle07} assigned half of
this  difference to non-thermal emission from a
beamed radio core in the quasars 
while the remaining half was thought
to arise  from MIR absorption in the radio galaxies. 
This combination agrees with standard AGN unification
schemes.

Except for cases borderline 
to flat spectrum quasars such as 3C\,245, our steep-spectrum objects lack such 
strong cores, and enhanced MIR emission 
due to non-thermal radiation seems unlikely. 
If the MIR continuum flux of the quasars in our sample were
 enhanced by synchrotron emission, they would lie
systematically at higher luminosities than the de-reddened radio galaxies 
(Fig.~\ref{fig4}). 
Furthermore, we would 
not expect the reddened quasars to fit the radio galaxies in both continuum level and 
silicate feature strength without any corrections for  non-thermal contributions 
(Fig.~\ref{fig1}). Thus, while a non-thermal component at some level
is not entirely 
ruled out, the results for our sample do not require one.

\subsection{Radio-normalized 15\,$\mu$m luminosity}
\label{sec:lowz}

Our data set  expands the comparison between low and
intermediate redshift (and luminosity) radio sources to  more
powerful sources. 
A comparison provides new insight on the luminosity dependence
of the radio-normalized 15\,$\mu$m dust emission.

\subsubsection{Comparison data}

\begin{figure}[t!]
\centering
\includegraphics[angle=0,scale=.46]{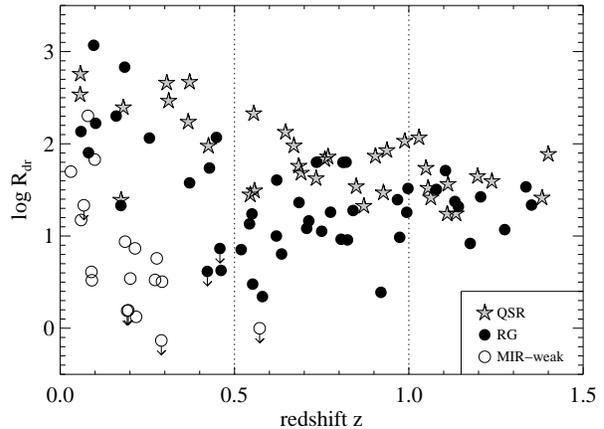}
\caption{$\log R_{\rm dr}$  
plotted versus the redshift $z$. $R_{\rm dr}$ is defined as $\nu L_{\nu,15\,\mu{\rm
      m}}/\nu L_{\nu,178\,{\rm MHz}}$, and 15~\micron\ and 178~MHz refer
to rest wavelength and frequency.\label{fig8}}
\end{figure}

We calculated
$R_{\rm dr}\equiv\nu
L_{\nu,15\,\mu{\rm m}}/\nu L_{\nu,178\,{\rm MHz}}$ for our objects
as well as for two comparison samples at $0.5<z<1.0$ and at $z<0.5$. For
the comparison  samples, we used all  FR-II \citep{fan74} radio
sources in the 3CRR catalog
observed spectroscopically with {\it Spitzer}. Spectroscopic MIR data
are available for 42 out of 49 sources (86\%) at $0.5<z<1.0$ and for
40 out of 60 sources  (67\%) at $z<0.5$. We used data from
\citet{ogl06}, \citet{cle07}, and \citet{har09} or retrieved the
spectra directly from the {\it Spitzer} archive and, after standard
data  reduction, determined the MIR fluxes.  Rest frame radio and MIR
luminosities were calculated  in the same manner as for our
high-redshift sample using observed spectral indices from the  3CRR
webpage.
Figure \ref{fig8} shows $R_{\rm dr}$ versus redshift.

Especially at lower redshift, a population of MIR-underluminous radio
galaxies has been found \citep[e.g.,][]{mei01,shi05,ogl06}. Such objects 
also have quite different multi-wavelength properties  compared to
MIR-luminous radio galaxies \citep[e.g.][]{but09,har09}, and they may 
represent a population of radio galaxies with low (radiative)
accretion power. We therefore present the statistics  below 
excluding all the ``MIR-weak'' objects with  
$\nu L_{\nu,15\,\mu{\rm m}} < 8.0 \times
10^{43}$\,erg/s \citep{ogl06}. This excludes one source at
$0.5<z<1.0$ and 17 sources at $z<0.5$. 
All three samples at the
different redshifts  include a small number of CSS sources.
Excluding those sources would not affect the statistics in any
significant way.

\begin{figure*}[t!]
\centering
\includegraphics[angle=0,scale=.5]{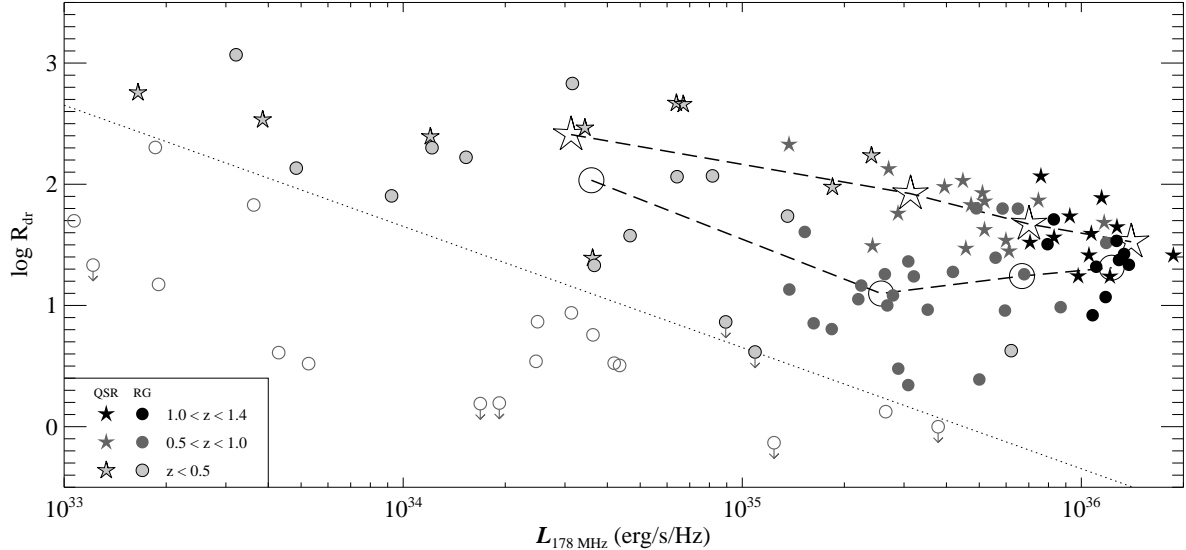}
\caption{$\log R_{\rm dr}$ 
($R_{\rm dr}\equiv\nu L_{\nu,15\,\mu{\rm m}}/\nu L_{\nu,178\,{\rm MHz}}$) 
plotted versus rest 178\,MHz luminosity for the three samples at $z<0.5$, $0.5<z<1.0$, and 
$1.0<z<1.4$. The dotted line indicates a MIR luminosity of 
$\nu L_{\nu,15\,\mu{\rm m}} = 8 \times 10^{43}$\,erg/s, which separates MIR-weak 
(small open symbols) and MIR-strong (filled symbols) sources \citep{ogl06}. 
The large open symbols show the mean values of $\log R_{\rm dr}$ and 
$L_{178\,{\rm MHz}}$ for the quasars and radio galaxies in four 
 $L_{178\,{\rm MHz}}$ luminosity bins (see Table~\ref{tab3}).
\label{fig6}}
\end{figure*}

\subsubsection{Luminosity dependence}

\begin{table*}
\begin{center}
\caption{\protect\centering Average  $R_{\rm dr}$ for quasars and radio galaxies in
 four  luminosity bins \label{tab3}}
\begin{tabular}{cr@{: }lcc}
\tableline\tableline
sample                                                         &
\multicolumn{2}{c}{number of} &
\multirow{2}{*}{$\langle \log R_{\rm dr} \rangle$} &
\multirow{2}{*}{$\frac{\langle R_{\rm dr}\rangle_{\rm QSRs}}{\langle R_{\rm dr}\rangle_{\rm RGs}}$}\\ 
       ($L_{178\,{\rm MHz}}$ in erg\,s$^{-1}$\,Hz$^{-1}$)    & \multicolumn{2}{c}{sources}  & & \\\hline
\multirow{2}{*}{$L_{178\,{\rm MHz}} < 1 \times 10^{35}$}                   & QSR & \hspace*{4pt}7 & $2.41 \pm 0.16$ & \multirow{2}{*}{$2.4 \pm 1.3$} \\
                                                                         &  RG & 11 & $2.03 \pm 0.18$ & \\\hline
\multirow{2}{*}{$1 \times 10^{35} < L_{178\,{\rm MHz}} < 5 \times 10^{35}$} & QSR & 10 & $1.92 \pm 0.09$ & \multirow{2}{*}{$6.6 \pm 1.9$} \\
                                                                         &  RG & 18 & $1.10 \pm 0.09$ & \\\hline
\multirow{2}{*}{$5 \times 10^{35} < L_{178\,{\rm MHz}} < 1 \times 10^{36}$} & QSR & 11 & $1.67 \pm 0.07$ & \multirow{2}{*}{$2.7 \pm 1.0$} \\
                                                                          & RG & 10 & $1.24 \pm 0.15$ & \\\hline
\multirow{2}{*}{$L_{178\,{\rm MHz}} > 1 \times 10^{36}$}                    & QSR & \hspace*{4pt}8 & $1.53 \pm 0.07$ & \multirow{2}{*}{$1.7 \pm 0.4$} \\
                                                                         &  RG & \hspace*{4pt}8 & $1.31 \pm 0.07$ & \\

\tableline
\end{tabular}

\tablecomments{The average $R_{\rm dr}$ is taken as 
$\langle R_{\rm dr}\rangle = 10^{\langle\log R_{\rm dr}\rangle}$ with  
$R_{\rm dr}\equiv\nu L_{\nu,15\,\mu{\rm m}}/\nu L_{\nu,178\,{\rm MHz}}$. MIR-weak sources 
with $\nu L_{\nu,15\,\mu{\rm m}} = 8 \times 10^{43}$\,erg/s have been excluded. 
Including the MIR-weak radio galaxies into the statistics does not change 
the 
trends: the $\langle\log R_{\rm dr}\rangle$ changes to 
$1.40 \pm 0.16$ and $0.94 \pm 0.12$ for the two lower luminosity bins, 
respectively. There are no MIR-weak objects in our sample for the two higher 
luminosity bins.}
\end{center}
\end{table*}

Figure~\ref{fig6} shows  $R_{\rm dr}$ as a function of radio luminosity
$L_{178\,{\rm MHz}}$. The most striking features of this figure are:

\begin{itemize}   
\item [1)] The $R_{\rm dr}$ of quasars decreases steadily with increasing
  $L_{178\,{\rm MHz}}$. This decline is seen  over almost three orders of 
  magnitude in $L_{178\,{\rm MHz}}$ and over the full
  redshift range (Fig.~\ref{fig8}) covered here.
  
\item [2)] 
  For the radio galaxies (excluding the MIR-weak sources), $R_{\rm
    dr}$ at low to intermediate $L_{178\,{\rm MHz}}$
  decreases similar to the quasars.  However, for 
  $L_{178\,{\rm MHz}}\ga 2\times10^{35}$\,erg/s/Hz, $R_{\rm dr}$ flattens
  or even increases towards the highest luminosities.  This
  leads to an apparent $R_{\rm dr}$ convergence of quasars and radio
  galaxies in the highest luminosity/redshift bin.

\end{itemize}  
In order to quantify these trends, we divided the objects into four 
luminosity bins in $L_{178\,{\rm MHz}}$. For each luminosity bin we then 
calculated the mean $\log R_{\rm dr}$ and the mean $L_{178\,{\rm MHz}}$ 
for the radio galaxies and the quasars separately. The results are 
indicated in Figure~\ref{fig6} and summarized in Table~\ref{tab3}.
These numbers confirm that the ratio of $R_{\rm dr}$ between 
radio galaxies and quasars changes with luminosity and that 
there seems to be a maximum difference at a luminosity $L_{178\,{\rm MHz}}$ 
of a few times $10^{35}$\,erg/s/Hz, where most
sources belong to the intermediate redshift ($0.5<z<1.0$) sample.
Because the quasars show a fairly continuous decrease in $R_{\rm dr}$
over all $L_{178\,{\rm MHz}}$ (or $z$) covered here, the variation in
the $R_{\rm dr}$ ratio is probably mostly due to the radio galaxies.

One possible explanation  of the trends we observe is that the 
fraction of the total power emitted by the radio lobes increases with 
$L_{178\,{\rm MHz}}$.  
Because for our flux-limited
sample an increase in radio luminosity on average translates  into an
increase in redshift, one could imagine increased lobe power due to
(on average) denser intergalactic environments at higher redshifts. 
Such effects, however, should apply  to radio galaxies and
quasars alike and cannot easily explain the $R_{\rm dr}$ upturn for
the most luminous radio galaxies.

The apparent $R_{\rm dr}$ decline  in Figure~\ref{fig6} could in principle  be
caused by the simple fact of plotting $1/L_{178\,{\rm MHz}}$
versus $L_{178\,{\rm MHz}}$, but this again would not
explain the rising $R_{\rm dr}$ for the most luminous radio
galaxies. Therefore we assume that there is a real physical effect
dependent on luminosity and differing for quasars and radio galaxies.

One example of a possible physical effect is that the dust covering angle may be
reduced  at high luminosities,  for example by destroying  dust or pushing it farther away
from the nucleus.  This would be consistent with the receding-torus model
\citep[e.g.][]{law91}.  A decline of $R_{\rm dr}$ with luminosity
is expected  in such a model if the height of the torus remains constant. 
However,  the height could be blown up by, for instance,
accompanying starbursts, and this could also be correlated with
luminosity \citep[][]{haas04}. But even in this picture the
behaviour in $R_{\rm dr}$  for the  most luminous radio galaxies remains a
puzzle.

An alternative explanation could involve the effects of additional
absorption of the radio galaxies' 15\,$\mu$m continuum emission by
dust in  the host galaxy, but outside the torus. A high dust column 
density in the host would be required for such behavior. In a situation 
where the dust
(including the torus) is pushed to larger  radii due to the increased
luminosity of the nucleus, the line-of-sight dust column density in
the host might become too small to make significant contributions to
the  total absorption of the continuum emission. Mid- and far-IR
observations of the optically-selected, mostly radio quiet
Palomar-Green (PG) quasars indicate that  the cold end of the dust
temperature distribution of the most luminous PG quasars at $1<z<2$ is
surprisingly warm compared with PG quasars of low and intermediate
luminosity \citep[][]{haas03}. This suggests that the amount of colder
dust in these (hyper)-luminous IR sources is relatively small and that
the AGN effectively heats much of the host galaxy's dust.  If this
scenario applies for our high-$z$ 3CRR radio galaxies as well, then
the total line-of-sight dust column density may be too small to be
able to absorb as much as   one would expect from a scaled-up version
of lower-luminosity AGNs, which are located in hosts relatively richer
in dust. In other words, if the AGN power relative to the host's dust
mass  exceeds a critical threshold, the maximum possible extinction
declines, hence the ratio of $R_{\rm dr}$(QSR)/$R_{\rm dr}$(RG)
declines as well.

Most likely there will not be a single explanation for the observed trends but 
a combination of effects will have to be considered. Additional insights may 
come from extending the luminosity range by including {\it Herschel} observations 
of the highest redshift 3CRR sources, or by including further constraints from 
a {\it Chandra} X-ray study of the present sample (B. Wilkes, in preparation).

\subsection{Emission Lines}

Four prominent emission lines can be identified in the average spectra
(Figure~\ref{fig1}): the low ionization [\ion{Ne}{2}] line, the
medium ionization species [\ion{S}{4}] and [\ion{Ne}{3}], and 
the high ionization AGN tracer [\ion{Ne}{5}].  We determined
emission-line fluxes only from the average spectra, but their uncertainties
still are large. This is for two reasons.  The individual
spectra in which emission lines can be identified show a wide range
of  emission-line equivalent widths for each type of 
object. This dispersion  among individual objects leads to considerable uncertainty for 
the line fluxes in the average spectra. In addition,  the 
measured emission-line fluxes in the average spectra depend strongly on the 
placement of the continuum.  Especially for weak or noisy features 
on top of a complex continuum, this results in  large errors. For
these reasons,
conclusions from the emission lines should be considered tentative.  
With these caveats in mind, within the uncertainties the
emission line ratios for radio galaxies and quasars are similar.
Moreover  the line luminosities 
are comparable as well
(Table~\ref{tab2}).

\begin{table}[t!]
\begin{center}
\caption{\protect\centering Emission-line luminosities.\label{tab2}}
\begin{tabular}{lcccc}
\tableline\tableline
           & [\ion{S}{4}] & [\ion{Ne}{2}] & [\ion{Ne}{5}] & [\ion{Ne}{3}]  \\
$\lambda$  & 10.5\,$\mu$m & 12.8\,$\mu$m  & 14.3\,$\mu$m  & 15.5\,$\mu$m   \\
\tableline
QSRs       & 2.5 & 1.2 & 1.9 & 4.6 \\
RGs        & 2.2 & 1.5 & 1.6 & 4.2 \\
\tableline
\end{tabular}
\tablecomments{The emission-line luminosities are measured in \\
the average spectra and given in $10^{43}$\,erg/s. The
 uncertainties \\are $\sim$25\% for the brightest line [\ion{Ne}{3}] 
and $\sim$50\% for the other\\ three species 
(see text).}
\end{center}
\end{table}

The spectra  cover the position of the 11.3\,$\mu$m PAH
feature, but this star formation tracer was not detected in the
average spectra. This
suggests that the MIR emission in our objects is strongly dominated by
the AGN and that any contributions from star formation are small
compared to AGN emission. 3CR radio sources at $z<1$ also  have less star 
formation than optically or NIR selected radio-quiet QSOs at comparable redshifts
\citep{shi07}. In fact, 
in most lower-redshift 3CR sources, PAH features remain undetected. This 
is consistent with the finding 
that many of the radio sources are
hosted by rather quiescent elliptical galaxies \citep[e.g.,][]{flo08}.

\section{Conclusions}

The 3CRR sample at $1 < z < 1.4$ represents the most radio-luminous 
steep-spectrum quasars and double-lobed radio galaxies 
for which the rest-frame MIR 9--16\,$\mu$m wavelength range is
accessible with the {\it Spitzer} Infrared Spectrograph. 
The sample is unbiased with respect to orientation. 

\begin{itemize}
\item[1)]
The quasar SEDs are very similar to each other in shape. They show 
a strong NIR and MIR continuum, silicates in emission, and
an emission bump at 2--5\,$\mu$m. 

\item[2)]
The mean radio galaxy spectrum ($8 < \lambda_{\rm rest} < 16$\,$\mu$m) is 
consistent with a quasar spectrum reddened by a dust screen of 
$A_{V}\sim20$\,mag both in spectral shape and in flux level.

\item[3)]
Radio galaxy extinctions indicated by their 2--5\,$\mu$m 
SEDs are higher than the extinction derived from the silicate
absorption and the (rest) 15\,$\mu$m flux. This can be explained if 
the component responsible for the NIR bump in the quasar SED 
suffers more extinction than the MIR emission.  

\item[4)]
All sources in our sample with silicate absorption have small core 
fraction ($<$$10^{-4}$), while objects with silicate emission often 
show stronger cores. This strongly supports the idea that the 
observed extinction in our sample is related to the orientation of 
the source.

\item[5)]
The average spectra show similar emission-line ratios for radio galaxies 
and quasars, and no PAH features are detected. This argues in favor of 
similar intrinsic AGN power and star-forming activity in quasars and radio 
galaxies, as predicted by  unified schemes. 

\item[6)]
The comparison of our high-$z$ sample with lower redshift 3CRR
sources reveals a steady decline of the quasar $R_{\rm dr}$ with increasing luminosity, 
and a partial decline, followed by a high-luminosity upturn, of the radio
galaxies' $R_{\rm dr}$. Possible explanations for these trends may involve 
an increased fraction of power emitted by the radio lobes due to different 
galactic environments at higher $z$ and/or different dust geometries due to the 
higher nuclear luminosities (e.g distance of the dust from the nucleus, solid 
angle of the dust exposed to the radiation, or total dust column density on the line 
of sight).


\end{itemize}
While the MIR spectra provide valuable insight into the nature 
of the MIR emission and the relation between 
the most powerful type-1 and type-2 radio AGN, 
we expect further constraints from  {\it Chandra} X-ray and 
{\it Herschel} far infrared/sub-mm observations of this sample.

\acknowledgments
{\noindent \it Acknowledgments:}
This work is based on observations made with the {\it Spitzer
Space Telescope}, which is operated by the Jet Propulsion Laboratory,
California Institute of Technology under a contract with NASA.
Support for this work was provided by NASA through an award issued by
JPL/Caltech.  This research has made use of the NASA/IPAC
Extragalactic Database (NED) which is operated by the Jet Propulsion
Laboratory, California Institute of Technology, under contract with
the National Aeronautics and Space Administration. 
M.H.\ is supported by Nordrhein-Westf\"alische Akademie der
Wissenschaften und der K\"unste.

{\it Facility:} \facility{{\it Spitzer} (IRS)}.

\clearpage

\clearpage

\end{document}